\documentclass[fleqn,usenatbib]{mnras}

\usepackage{times}

\usepackage[T1]{fontenc}
\usepackage{ae,aecompl}

\usepackage{graphicx}	
\usepackage{amsmath}	
\usepackage{amssymb}	
\usepackage{rotating}
\usepackage{tikz}
\hypersetup{pdfauthor={A. P. Rushton et al.}, pdftitle={Disk-Jet quenching of the Galactic Black Hole Swift\,J1753.5-0127}}

\newcommand{\degree}{\hbox{$^\circ$}}

\def \vhel{\ifmmode{~V_{{\rm HEL}}}\else{~$V_{{\rm HEL}}$}\fi}
\def \vsys{\ifmmode{~V_{{\rm SYS}}}\else{~$V_{{\rm SYS}}$}\fi}
\def \HA {\ifmmode{{\rm\H}\alpha}\else{${\rm\ H}\alpha$}\fi}

\def \msun{\ifmmode{{\rm\ M}_\odot}\else{${\rm\ M}_\odot$}\fi}

\def \myr{\ifmmode{{\rm\ M}_\odot{\rm\ yr}^{-1}}
        \else{${\rm\ M}_\odot$ yr$^{-1}$}\fi}

\def \mdot{\ifmmode{\dot{M}}\else{$\dot{M}$}\fi}
\def \tena#1 #2 {\ifmmode{#1 \times 10^{#2}}\else{$#1 \times 10^{#2}$}\fi}
\def \kms{\ifmmode{~{\rm km\,s}^{-1}}\else{~km s$^{-1}$}\fi}

\def \apj{ApJ}
\def \mnras{MNRAS}

\def \aap{A\&A}

\def \apjs{ApJS}
\def \apjl{ApJL}

\title[Disk-Jet quenching of a Galactic black hole]{Disk-Jet quenching of the Galactic Black Hole Swift\,J1753.5-0127}

\author[A. P. Rushton et al.]{A. P. Rushton,$^{1,\,2}$\thanks{E-mail: Anthony.Rushton@physics.ox.ac.uk (APR)}
A. W. Shaw,$^{2}$
R. P. Fender,$^{1}$
D. Altamirano,$^{2}$
P. Gandhi,$^{2}$
\newauthor
P. Uttley,$^{3}$
P. A. Charles,$^{2}$
M. Kolehmainen,$^{4,\,1}$
G. E. Anderson,$^{5,\,1}$
\newauthor
C. Rumsey$^{6}$
and
D. J. Titterington$^{6}$
\\
$^{1}$Department of Physics, Astrophysics, University of Oxford, Keble Road, Oxford OX1 3RH, UK\\
$^{2}$School of Physics and Astronomy, University of Southampton, Highfield, Southampton SO17 1BJ, UK\\
$^{3}$Anton Pannekoek Institute, University of Amsterdam, Science Park 904, NL-1098 XH Amsterdam, the Netherlands\\
$^{4}$Observatoire Astronomique de Strasbourg, 11 Rue de l'Universit\'{e}, 67000 Strasbourg, France\\
$^{5}$International Centre for Radio Astronomy Research, Curtin University, GPO Box U1987, Perth, WA 6845, Australia\\
$^{6}$Astrophysics Group, Cavendish Laboratory, 19 J. J. Thomson Avenue, Cambridge CB3 0HE, UK
}

\date{Accepted XXX. Received YYY; in original form ZZZ}

\pubyear{2015}

\begin{document}
\label{firstpage}
\pagerange{\pageref{firstpage}--\pageref{lastpage}}
\maketitle

\begin{abstract}
We report on radio and X-ray monitoring observations of the BHC Swift\,J1753.5-0127 taken over a $\sim10$ year period. Presented are daily radio observations at 15 GHz with the AMI-LA and X-ray data from {\it Swift} XRT and BAT. Also presented is a deep 2hr JVLA observation taken in an unusually low-luminosity soft-state (with a low disk temperature).  We show that although the source has remained relatively radio-quiet compared to XRBs with a similar X-ray luminosity in the hard-state, the power-law relationship scales as $\zeta=0.96\pm0.06$ i.e. slightly closer to what has been considered for radiatively inefficient accretion disks. We also place the most stringent limit to date on the radio-jet quenching in an XRB soft-state, showing the connection of the jet quenching to the X-ray power-law component; the radio flux in the soft-state was found to be $<21~\mu$Jy, which is a quenching factor of $\gtrsim25$.
\end{abstract}




\section{Introduction}

Galactic X-ray Binaries (XRBs) are a powerful tool for studying the production of relativistic jets from accretion disks. XRBs exhibit distinct spectral states over observable time-scales, allowing the study of the disk-jet relationship with vastly different accretion rates and disk topologies \citep[reviewed by][]{2010LNP...794..115F,2015ASSL..414...25G}. Typical ``hard-states'' are characterised by a hard X-ray power-law spectrum ($\Gamma \sim1.5$), with strong variability ($\sim20-50\%$~rms) and a compact radio counterpart. At high accretion rates, XRBs can transition to ``soft-states'', with the spectrum becoming dominated by a multi-temperature blackbody component and a steepening of the hard power-law component; the compact radio counterpart becomes quenched in the soft-state and some sources release discrete knots of ejecta during the hard-to-soft transition. 

Swift\,J1753.5-0127 (hereafter Swift\,J1753.5) was discovered by the Burst Alert Telescope \citep[BAT;][]{2005SSRv..120..143B} in 2005 \citep{2005ATel..546....1P} as a hard-spectrum ($\gamma$-ray source) transient at a relatively high Galactic latitude (+12\degree). The source luminosity peaked within a week, at a flux of $\sim200$ mCrab, as observed by the Rossi X-Ray Timing Explorer ({\it RXTE}) All Sky Monitor (ASM; 2--12 keV) \citep{2007ApJ...659..549C}. The source was also detected in the UV, with the Ultraviolet/Optical Telescope \citep[UVOT;][]{2005ATel..553....1S}, and in the radio with MERLIN \citep{2005ATel..558....1F}. An R $\sim15.8$ optical counterpart was identified by \citet{2005ATel..549....1H}, who noted that it had brightened by at least 5 magnitudes (as it is not visible in the Digitized Sky Survey; DSS), thereby establishing Swift\,J1753.5 as a Low Mass X-ray Binary (LMXB) with a very faint, low-mass donor. Subsequent time-resolved photometry of the optical counterpart revealed R-band modulations on a period of 3.2h, which are indicative of the orbital period ($P_{orb}$) of the system \citep{2008ApJ...681.1458Z}.

Almost immediately after its peak the X-ray flux of Swift\,J1753.5 started declining, but after $\sim100$~days it remained roughly constant at $\sim20$ mCrab (2--12 keV) for over 6 months rather than returning to quiescence as might have been expected for a typical Black Hole X-ray Transient BHXRT \citep{2006csxs.book..215C}. The source has still not returned to quiescence, $\sim10$ years after its initial discovery, and has instead exhibited significant long-term ($> 400$~day) variability over the course of its prolonged `outburst' \citep{2013MNRAS.433..740S}. Swift\,J1753.5 has remained as a persistent LMXB in a hard accretion state for the majority of this time, however it has experienced a number of short-term spectral softenings, characterised by an increase in the temperature of the inner accretion disk and simultaneous steepening of the power law component in the X-ray spectrum \citep{2015PASJ...67...11Y}. Investigation of the source during one such event with {\it RXTE} revealed that it had transitioned to a hard intermediate accretion state. However, unlike the majority of BHXRTs, Swift\,J1753.5 did not continue towards an accretion disk dominated soft-state and instead returned to the hard-state \citep{2013MNRAS.429.1244S}. The durations of these `failed state transitions' have typically been short ($\sim25$~d), but in early 2015, the source appeared to undergo another state transition when the {\it Swift}-BAT flux appeared to drop to its lowest levels since the source's discovery \citep{2015ATel.7196....1O}. Subsequent follow-up with the {\it Swift} X-ray Telescope \citep[XRT;][]{2005SSRv..120..165B}, {\it XMM-Newton} \citep{2001A&A...365L...1J} and the Nuclear Spectroscopic Telescope Array \citep[{\it NuSTAR};][]{2013ApJ...770..103H} revealed that Swift\,J1753.5 had transitioned to one of the lowest luminosity soft-states recorded in LMXBs \citep{2015ATel.7216....1S,2016MNRAS.458.1636S}.

With a large ($\Delta$R $\sim$5 mag.) optical increase at outburst, we would not expect to detect any spectroscopic signatures of the donor, due to the optical light being dominated by the accretion disk. \citet{2009MNRAS.392..309D} confirmed this with spectroscopic observations revealing a smooth optical continuum and no evidence for features associated with the donor. With no detectable fluorescence emission either, it has not been possible to obtain any direct evidence of the compact object mass. However, INTEGRAL observations highlighted the presence of a hard power-law tail up to $\sim600$~keV, very typical of a black hole candidate (BHC) in the hard-state \citep{2007ApJ...659..549C}. Also, the power density spectrum from a pointed {\it RXTE} observation revealed a 0.6 Hz quasi-periodic oscillation (QPO) with characteristics typical of BHCs \citep{2005ATel..550....1M}. QPOs have also been seen at 0.08 Hz in optical data \citep{2009MNRAS.392..309D} as well as in a number of X-ray observations after the initial outburst had declined \citep{2007MNRAS.378..182R,2007ApJ...659..549C}.

Recently, \citet{2014MNRAS.445.2424N} reported on evidence for a low mass (< 5\msun) BH in Swift\,J1753.5, based on observations of narrow features in the optical spectrum which they associate with the donor, despite such features not being identifiable or visible in previous spectroscopic studies \citep{2009MNRAS.392..309D}. Given the high Galactic latitude of Swift\,J1753.5, \citet{2007ApJ...659..549C} concluded that its distance is likely 4--8 kpc. However, in recent work fitting the UV spectrum with an accretion disk  model and assuming a 5\msun BH, \citet{2014ApJ...780...48F} obtain a distance of $< 2.8$~kpc and $<~3.7$~kpc for a binary inclination of i = $55\degree$ and $0\degree$, respectively.

\section{Observations}

Since the initial discovery and outburst in May/June of 2005 \citep{2005ATel..546....1P}, the source has mostly remained in the low hard-state (with the exception of the aforementioned failed state transitions); however, towards the start of 2015 Swift\,J1753.5 appeared to enter an unusual low luminosity soft-state \citep{2015ATel.7196....1O}. Estimates for the unabsorbed luminosity in the 0.7--78 keV band was found to be only $\approx0.6 \%$ of the Eddington luminosity, even though the source clearly transitioned to the soft-state \citep{2016MNRAS.458.1636S}. In this paper we analyse all the radio monitoring observations of this source since its initial discovery and report deep limits on the radio activity in the soft-state.

\subsection{Archival radio observations (2005--2009)}

Following the initial discovery of Swift\,J1753.5 in 2005 an intensive series of radio observations was reported by \citet{2010MNRAS.406.1471S}. Their work analysed a sub-set of observations that were coincidental with {\it Swift} and {\it RXTE}. Here we make use of the full set of radio fluxes which they reported in table A2, including the VLA at 1.4, 4.8 and 8.4 GHz, MERLIN at 1.7 GHz and WSRT 4.9 and 8.5 GHz. The initial observations caught a radio flare that peaked up to $\sim3$~mJy and decayed over a period of about 100 days, until steadying at a mean flux of $\sim0.4$~mJy. The radio spectrum between 1.4 and 8.4 GHz showed a mostly flat or (slight) suggestion of inverted spectrum, indicative of a self-absorbed synchrotron jet of energetic particles. The overall radio flare, decay and steady emission closely couple with the fluxes seen in the X-ray band.

\subsection{AMI-LA monitoring (2013--2015)}

Between January 2013 and July 2015 the Arcminute Microkelvin Imager Large Array (AMI-LA) aperiodically monitored Swift\,J1753.5, with a typical cadence of 1--2 weeks. Each epoch lasted a few hours and was scheduled around local transit of the telescope. The array consisted of eight 12.8 m dishes with baselines ranging from 18 to 110 m located in Cambridge, UK and was built by the Mullard Radio Astronomy Observatory \citep[AMI Consortium:][]{2008MNRAS.391.1545Z}. During this period AMI-LA operated at a frequency range of 13.9--17.5 GHz, with the analogue XF correlator providing five usable channels each with a bandpass of 0.72 GHz giving a total bandwidth of 3.6 GHz.

Data calibration was performed using the \textsc{python} \textsc{drive-ami} pipeline \citep{2015A&C....13...38S,2015ascl.soft04005S}\footnote{\url{https://github.com/timstaley/drive-ami}}, which uses the MRAO tool \textsc{reduce} to automatically flag for interference, shadowing and hardware errors, conduct phase and amplitude calibrations, and Fourier transforms the data into \textit{uv}-FITS format \citep[see][]{2013MNRAS.429.3330P}. Imaging was performed within the Common Astronomy Software Application \citep[\textsc{CASA};][]{2007ASPC..376..127M} produced by NRAO, using the \textsc{clean} task driven by the \textsc{drive-ami} wrapper. The resultant maps were then loaded and analysed by the Transient Pipeline \cite[Trap:][]{2015A&C....11...25S} in order to extract the flux density of each epoch.

\subsection{New JVLA observation in May 2015}

Swift\,J1753.5 was observed with the Karl G. Jansky Very Large Array (JVLA) on 2015 May 13 07:32--09:15 UT (project code 15A-481), with a total on-source time of 84 minutes. The array was mostly in the ``B" configuration with baselines of up to 11.1 km (a single antenna had been moved to ``A'' configuration). The wide X-band receiver system was used with 3-bit samplers, allowing a tuneable frequency range of 7.7--12.6~GHz. Due to RFI within the band, the basebands were tuned to cover 7.98--11.67~GHz giving a total usable bandwidth of 3.69 GHz. Observations were made at a central frequency of 9.8 GHz using two basebands each divided into 16 sub-bands with 64 x 2 MHz channels per sub-band. 
 
Data reduction was carried out using \textsc{CASA}. Initial inspection of the data was performed to flag bad data from antenna errors, shadowing, RFI, or other instrumental issues. Hanning smoothing was applied to minimise the effects of RFI on surrounding frequency channels. Bandpass and flux scale calibration was performed using a short observation of 3C 286. The flux density scale was set using \citet{2013ApJS..204...19P} and transferred to the phase calibrator and target field. Time-dependent amplitude and phase gains were solved using the nearby phase calibrator J1743-0350 using a 2:8 minute duty cycle between calibrator and target. 

No significant radio flux was detected within the primary beam. The mean RMS noise towards the phase centre was 7~$\mu$Jy~bm$^{-1}$, making the $3\sigma$ upper limit of the flux from this epoch $<21~\mu$Jy.

\begin{figure}
	\includegraphics[width=8cm]{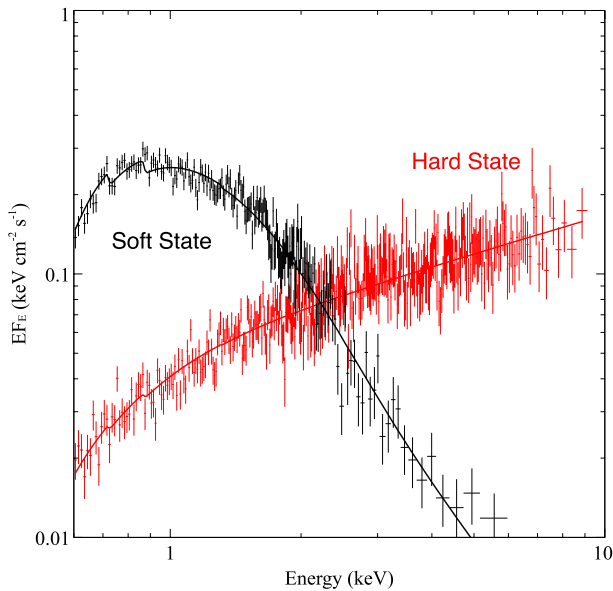}
    \caption{A comparison of unfolded {\it Swift}-XRT spectra from two epochs: during the 2015 soft-state (2015 May 13th; black) and during a typical hard-state (2013 June 18; red). The 2015 data are fitted with DISKBB+POWERLAW and the 2013 data are fitted with POWERLAW (solid lines). Adapted from \protect\citet{2016MNRAS.458.1636S}.}
    \label{fig:XRT_spectra}
\end{figure}

\begin{figure*}
	\includegraphics[width=19cm]{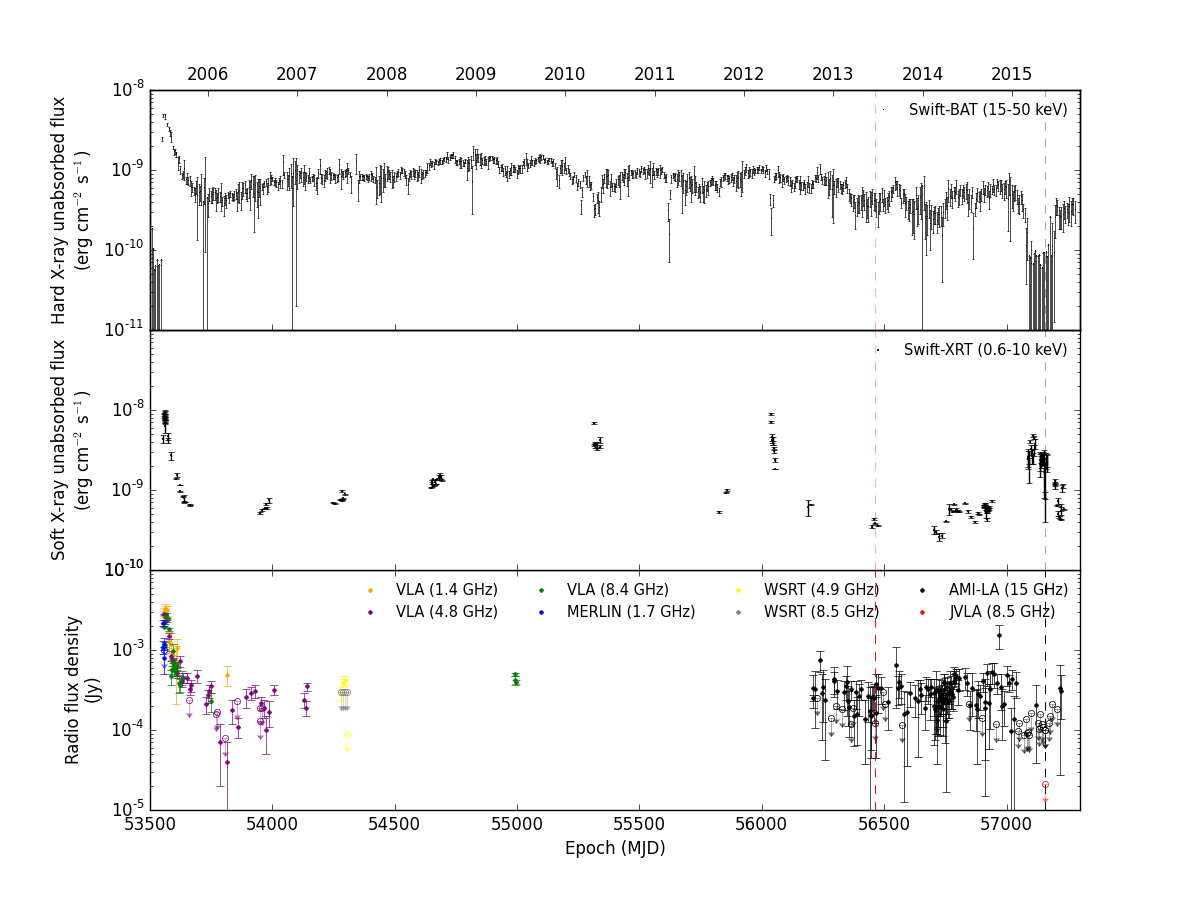}
    \caption{The 2005--2015 X-ray and radio lightcurve of Swift\,J1753.5. The X-ray spectra at times of the dotted lines are shown in Figure~\ref{fig:XRT_spectra}.}
    \label{fig:lightucrve}
\end{figure*}

\subsection{{\it Swift} observations}

We utilised 175 observations of Swift\,J1753.5 made with the {\it Swift} Gamma-ray Burst Mission \citep{2004ApJ...611.1005G} between 2005 July 02 and 2015 July 30. For the purposes of this work we have used a number of archival XRT observations (Target IDs: 00030090, 00031232 and 00033140), and we have also monitored the 15--50 keV flux with the BAT.

{\it Swift}-XRT operated in windowed timing (WT) mode for each of these observations, eliminating the possibility of photon pile-up on the CCD. Data were reduced using the \textsc{heasoft} v6.16 task \textsc{xrtpipeline} and count rates extracted from a circular region 20 pixels in radius ($\approx$ 47''). The background count rate was extracted from an annulus centred on the source with inner and outer radii of 80 and 120 pixels, respectively. Spectra were grouped to have a minimum of 20 counts per energy bin, allowing us to use the $\chi^2$ statistic when fitting models to the data.

All spectral fits were performed using \textsc{xspec} v12.8.2 \citep{1996ASPC..101...17A} which uses the $\chi^2$ minimisation technique to determine the best fit model. The interstellar absorption is accounted for by the \textsc{tbabs} model with \citet{2000ApJ...542..914W} abundances and photo-ionisation cross-sections described by \citet{1996ApJ...465..487V}. We also included a systematic error of 3\% for the {\it Swift}-XRT spectra, given the uncertainties of the response matrix (\textsc{swift-xrt-caldb-09}\footnote{\url{http://www.swift.ac.uk/analysis/xrt/files/SWIFT-XRT-CALDB- 09_v16.pdf}}). We fitted two models to each spectrum, an absorbed power-law (\textsc{powerlaw}) and an absorbed disk-blackbody plus power-law (\textsc{diskbb+powerlaw}), with the best-fit model determined using an F-test. Once this had been done, we extracted the unabsorbed flux with the \textsc{xpsec} model \textsc{cflux} and determined the 90\% confidence intervals for each fit parameter.

A {\it Swift}-XRT observation of Swift\,J1753.5 was performed on 2015 May 13 10:15--10:33 UT, 1~hr after the JVLA observation. The total exposure time was 1058s. The spectrum was extracted using similar methods to those described above and is well constrained with an absorbed disk-blackbody plus power-law. We measure the unabsorbed X-ray flux (0.6-10 keV) to be $F = 1.09^{+0.27}_{-0.37}\times10^{-9}$~erg~cm$^{-2}$~s$^{-1}$ with a photon index of $\Gamma=4.29^{+0.28}_{-0.71}$, an inner disk temperature $kT_{in}=0.31^{+0.05}_{-0.03}$~keV and $N_H=4.1^{+0.9}_{-1.8}\times10^{21}$~cm$^{-2}$ ($\chi^2/\rm{d.o.f.}=1.02$). The soft-state spectrum is shown as black in Figure~\ref{fig:XRT_spectra} and compared to a typical hard-state spectrum in red.

\section{Results and Analysis}

\subsection{Radio and X-ray lightcurve over $\sim10$ years}

The lightcurve shown in Figure~\ref{fig:lightucrve} initially shows the outburst of 2005, with an increase in both X-ray and radio luminosity by over an order of magnitude compared to the subsequent long-term average. The X-ray emission flared to almost $10^{-8}$~erg~cm$^{-2}$~s$^{-1}$ in both the 1-10 keV and 15-150 keV band with a slow decay over $\sim100$ days. Likewise the radio emission showed a quick rise to just over 3 mJy followed by a slow decay to around 0.4 mJy. 

There appeared to be two so-called ``failed state-transitions" \citep{2013MNRAS.429.1244S,2015PASJ...67...11Y} that occurred around MJD 55400 and 56000 that showed the source to increase in soft flux but no significant increase was seen at harder bands. Unfortunately no radio observations were available during those epochs.

Between 2013 and mid-2015 the AMI-LA telescope intensively monitored the source every few days. During the hard state, the flux maintain a time-averaged value of 0.3~mJy with an rms-variability (after correcting for noise) of 0.14~mJy; this behaviour was typical for the level post-2005 flare, assuming there was no major change to the spectral index. The source remained in this state until January 2015 when the flux suddenly quenched below the $3\sigma$ monitoring detection limit ($<150~\mu$Jy bm$^{-1}$). This corresponded to a soft-state transition seen in the X-rays; the hard X-ray emission dropped by about an order of magnitude and the soft X-rays slightly increased \citep[see][for a detailed description of this low luminosity soft-state] {2016MNRAS.458.1636S}. Swift\,J1753.5 remained in the radio-quenched soft-state for $\sim170$~days before returning to the hard-state in July of 2015. Two weeks later we stopped monitoring with AMI-LA due to a scheduled correlator upgrade.

\subsection{The X-ray / Radio Correlation of Swift\,J1753.5}
\label{sec:correlation}
\begin{figure}
	\includegraphics[width=9cm]{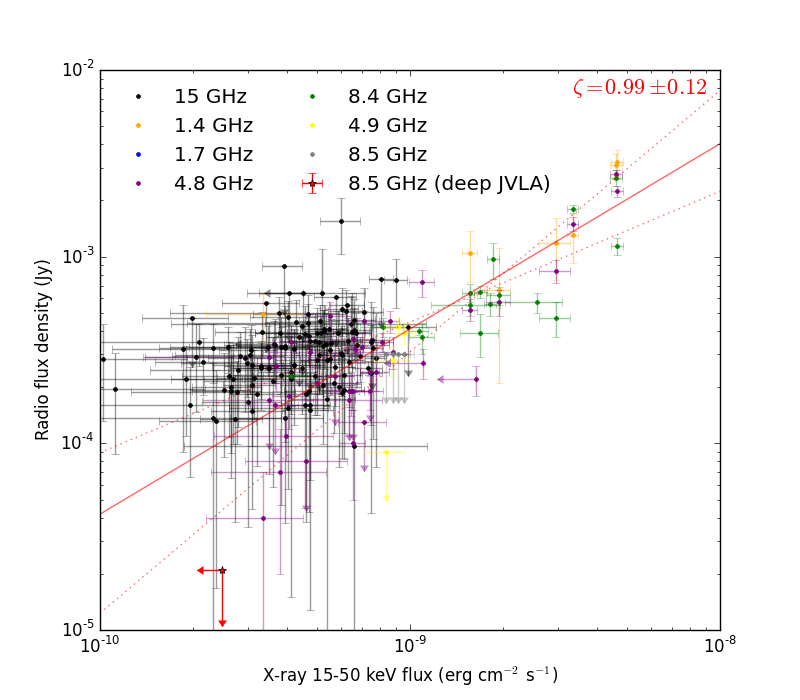}
	\includegraphics[width=9cm]{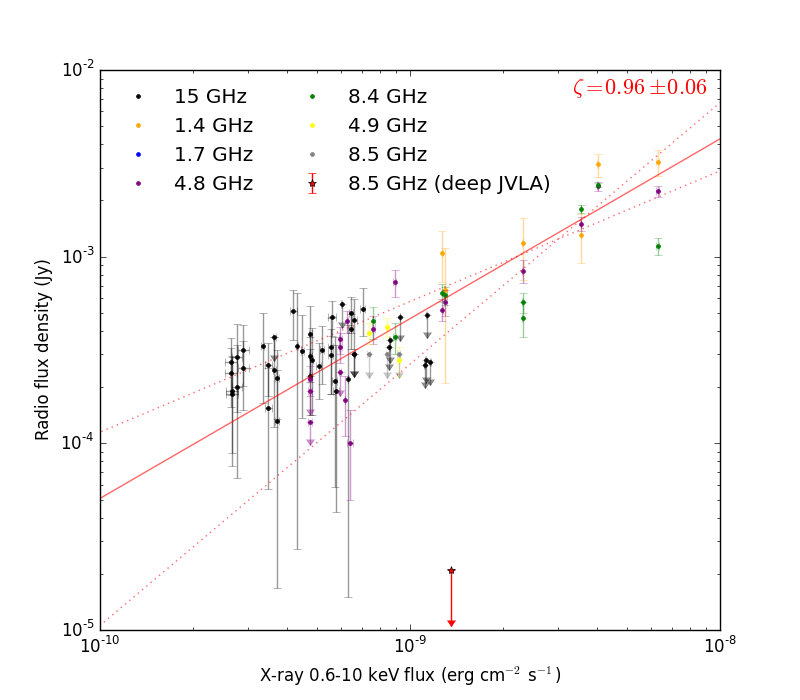}
	    \caption{The X-ray / radio flux correlation. The dotted lines show a powerlaw relationship of $\zeta=0.7$ and 1.4 and the solid line shows the best fit. \textit{Upper panel} shows the sample correlated with daily {\it Swift}-BAT data. \textit{Lower panel} shows the radio sample correlated with measurements taken with the {\it Swift}-XRT. Marked in both panels is the limit of a deep JVLA observation taken in the soft-state.}
    \label{fig:flux_correlation}
\end{figure}

\begin{figure*}
	\includegraphics[width=19cm]{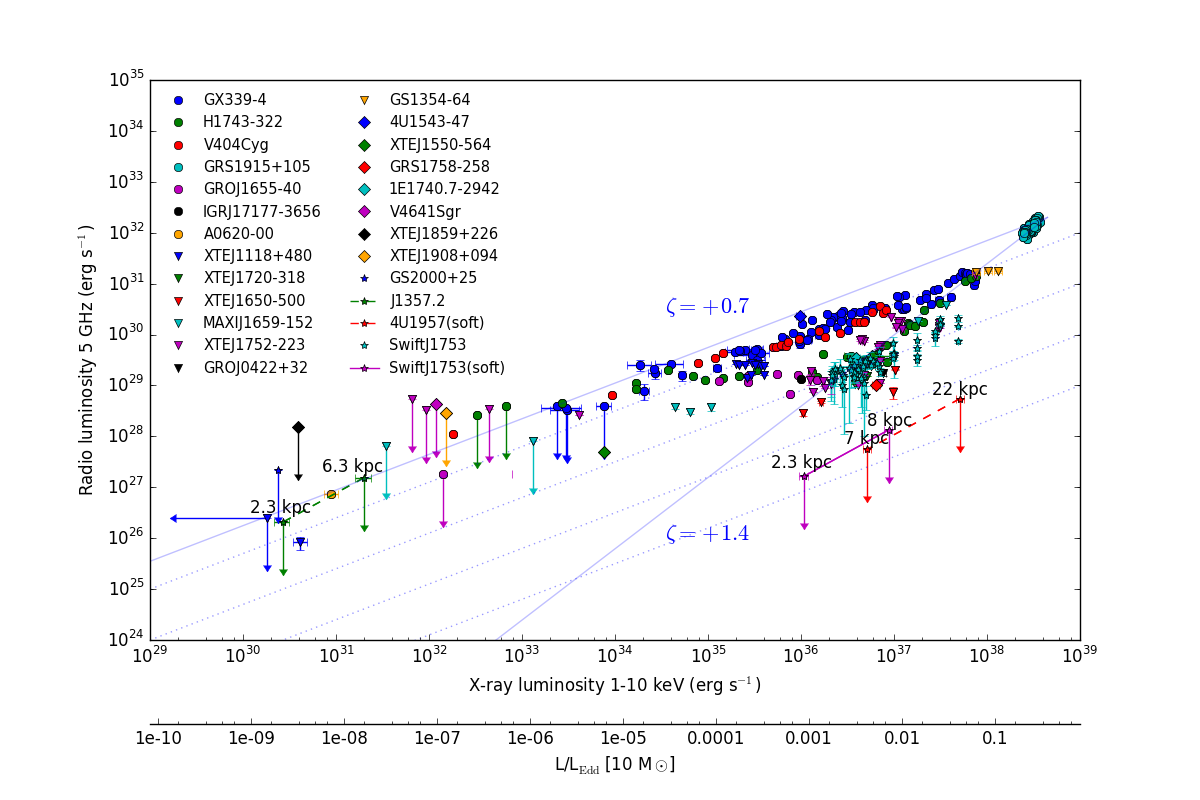}
    \caption{The X-ray/radio luminosity relationship for 25 XRBs. The solid lines show two different powerlaw relationships scaled from the brightest known XRB GRS1915+105. Also shown as dotted lines is the ``standard'' powerlaw relationship ($\zeta=0.7$) normalised by four different decades. For detections of Swift\,J1753 we assume a distance of 8 kpc.}
    \label{fig:luminosity}
\end{figure*}

We correlated the radio and X-ray flux over the entire $\sim10$ yr dataset by matching epochs that are quasi-contemporaneous. A maximum separation limit of three days was used to produce a reasonably large sample that did not significantly increase $\chi^2$ in our fits. All observations correspond to the hard or quiescent state (with exception of the deep JVLA soft-state epoch reported here, which is omitted from our analysis). We omitted the 1.7 GHz observations reported by \citet{2010MNRAS.406.1471S} as the MERLIN measurements are inconsistent with the overall spectrum and appear unusually low, which could be due to calibration errors (reanalysis of this data are beyond the scope of this paper).

The correlation was performed using data from both the soft 0.6--10 keV X-ray band measured by {\it Swift}-XRT and the hard 15--50 keV X-ray band measured by {\it Swift}-BAT. Figure~\ref{fig:flux_correlation} shows the radio relationship against the hard and soft bands in the upper and lower panels respectively. The majority of the lower radio luminosity data was taken using the AMI telescope which had a typical RMS noise of $\sim50~\mu$Jy~bm$^{-1}$ and most of AMI-LA (15 GHz) points are 2--10$\sigma$; we thus set a significance cut-off of 3$\sigma$ and place an upper-limit on any points below this flux. We performed an F-test by using an orthogonal distance regression analysis (using \textsc{scipy.odr}) with a single and broken powerlaw function to test if either model is favoured. The F-test showed that applying a broken power law was not justified and in Figure~\ref{fig:flux_correlation} show the powerlaw fit (solid red line) of the form $S_{\rm{radio}}=k(S_{\rm{X-ray}})^{\zeta}$, where $\zeta=0.99\pm0.12$ for the BAT and $\zeta=0.96\pm0.06$ for the XRT. 

Although estimating the disk-jet luminosity is dominated by uncertainty in the distance, it is clear that compared to other XRBs Swift\,J1753.5 has remained on the radio-quiet branch \citep[as originally shown in][]{2010MNRAS.406.1471S}. Assuming a distance of 2.3--8~kpc the median X-ray luminosity is $(5-50)\times10^{35}$~erg~s$^{-1}$ and is associated with a (5~GHz) radio luminosity of $(1-10)\times10^{28}$~erg~s$^{-1}$ i.e. an $L_{\rm{radio}}/L_{\rm{X-ray}}$ ratio of $\sim7.5$ dex. For most XRBs that follow the ``upper-branch'' of the scaling relationship (i.e. scale with powerlaw $\zeta\sim0.7$) the corresponding radio luminosity would be $(6-30)\times10^{29}$~erg~s$^{-1}$ i.e. a ratio of $\sim6$~dex. 

In Figure~\ref{fig:luminosity} we show the 5~GHz vs 1.6--10 keV luminosity of Swift\,J1753.5 (assuming the upper distance of 8 kpc) compared to a sample of quiescent and hard-state XRBs taken from \citet{2013MNRAS.428.2500C}. Swift\,J1753.5 falls on the ``lower-branch" which has a cluster of 4-5 sources that are also about an order of magnitude quieter in radio emission than e.g. GX\,339-4 and V404\,Cyg (the ``upper-branch"). 

\subsection{Soft-state quenching of the radio jet}

Our deep JVLA radio observation of Swift\,J1753.5 places the most stringent quenching factor on the soft-state radio flux from a transient BHC XRB. Previous works studying the radio quenching have only used the apparent XRB Radio-X-ray luminosity relationship to estimate the analogous hard-state radio flux. For example, \citet{2011ApJ...739L..19R} have so-far placed the deepest constraints on the radio flux of a persistent soft-state BH XRB; the soft-state radio flux density of 4U1957+11 was measured to be significantly less ($3\sigma$) than $11.4~\mu$Jy ($F_{\rm{5~GHz}}<5.7\times10^{-19}$~erg~s$^{-1}$~cm$^{-2}$) and the corresponding unabsorbed X-ray flux was measured as $F_{\rm{1-10~keV}}= 8.8\times10^{-10}$~erg~s$^{-1}$~cm$^{-2}$, thus making $\frac{F_{\rm{1-10~keV}}}{F_{\rm{5~GHz}}} > 1.5 \times 10^{9}$. The \textit{inferred} hard-state radio flux is between $3.7-9.2$~mJy, based on the assumption the source traces the upper-radio branch of the Radio-X-ray luminosity relationship and a distance of 7--22~kpc. If correct, the resultant radio quenching factor would be >2.5 orders of magnitude, implying that jets in the soft-state can be utterly suppressed.

In this paper we place a comparable tight constraint on the soft-state upper radio flux limit for a source with a \textit{known hard-state radio flux}. Swift\,J1753.5's soft-state radio flux was found to be significantly less ($3\sigma$) than $<21 \mu$Jy ($F_{\rm{5~GHz}}<1.05\times10^{-18}$~erg~s$^{-1}$~cm$^{-2}$) and the unabsorbed soft X-ray flux for this epoch was $F_{\rm{1-10~keV}}= 3.37\times10^{-10}$~erg~s$^{-1}$~cm$^{-2}$. Thus the soft-state radio to X-ray flux ratio of $\frac{F_{\rm{1-10~keV}}}{F_{\rm{5~GHz}}}>3.2\times10^{8}$ is within an order of magnitude to 4U\,1957+11. Scaling the X-ray to radio flux (using the empirical flux relationship found in figure~\ref{fig:flux_correlation}) corresponds to a hard-state flux of $F_{\rm{radio}}\approx600~\mu$Jy (N.B. the mean hard-state flux is $\sim300~\mu$Jy). Thus we find a direct quenching factor of >25.

We show the radio luminosity limits of Swift\,J1753.5 and 4U\,1957+11 in Figure~\ref{fig:luminosity}. Given the estimated distance to the two sources, the soft-state quenching factor relative to the upper-branch is approximately the same (within the uncertainty of the distance). Shown as a solid blue line is the powerlaw scaling of $\zeta=0.7$ and $1.4$ from GRS1915+105 (i.e. a 10 \msun~BH accreting at $\sim0.1~L/L_{\rm{Edd}}$). The soft-state jet is shown to quench at least an order of magnitude more than the so-called ``radiative-efficient" branch of GRS1915+105 \citep{2010A&A...524A..29R}, if it were accreting at the $\sim0.01~L/L_{\rm{Edd}}$ rate of Swift\,J1753.5 and 4U\,1957+11.

It is important to note that Swift\,J1753.5 is a more `canonical'  transient XRB than 4U\,1957+11 as the latter is a persistent soft X-ray emitter whilst the former a typical XRB transient that has transitioned across different X-ray states \citep[although the soft-state for Swift\,J1753.5 is unusually low -- see][]{2016MNRAS.458.1636S}. Also the radio emission from Swift\,J1753.5 is clearly under-luminous compared to typical hard-state XRBs (see section~\ref{sec:correlation}) and the quenching factor with respect to the upper-branch of the empirical relationship is similar to 4U\,1957+11. Moreover, the radio emission from 4U\,1957+11 has never been detected, thus we do not yet know the typical radio flux (assuming the source could ever produce a radio emitting jet).

\section{Discussion and Conclusions}

XRBs have a well-established, albeit complex phenomenological relationship between the state of the accretion disk and the jet luminosity. To first-order approximation there appears to be a correlation with the hard X-rays and the compact jet. \citet{2004MNRAS.355.1105F} showed that as XRBs trace a hysteresis around the Hardness Intensity Diagram (HID), they display distinctly different modes of jet activity; hard-states are associated with compact jets, soft-states with jet suppression and transitions from hard to soft-states can eject knots of discrete plasma. Intermediate states, such as those seen in Cygnus X-1 \citep{2012MNRAS.419.3194R}, have also shown a reduction in the jet luminosity and size (although not a complete quenching). Furthermore, within the hard-state it has been shown that a non-linear relationship exists between the luminosity of the jet and the accretion disk. For example, \citet{2003A&A...400.1007C} initially showed that for the BH XRB GX 339-4 the radio emission scales with the X-rays with a powerlaw index of $\zeta\sim0.7$. Then, by accounting for distance, it was shown that a sample of BH XRBs follows a similar luminosity relationship \cite[e.g.][]{2003MNRAS.344...60G,2006MNRAS.370.1351G}. Moreover, it has been suggested this relationship is a ``fundamental-plane" for black holes of all mass taking the form $L_r=(0.6\pm0.1) \log L_x +(0.8 \pm0.1) \log M + (7.3\pm4.1)$ \citep{2003MNRAS.345.1057M}.

Although evidence of a disk-jet relationship is ubiquitous, significant scatter in the X-ray/radio relationship has also been found. Some BHCs have shown a scaling relationship much closer to that proposed for radiatively efficient neutron stars \citep{2006MNRAS.369.1451K}. GRS1915+105 is known to have a much more luminous radio to X-ray ratio than other XRBs and \citet{2010A&A...524A..29R} found the variability in the hard (``plateau'') state to be significantly steeper, with a powerlaw index of $\zeta=1.7\pm0.3$; Swift\,J1753.5 on the other hand is significantly radio-quiet compared to the ``standard" GX~339-4 relationship. Evidence for a transition in the powerlaw relationship (i.e. from the radio quiet to loud track) has been found for H\,1743-322, which showed a possible transition from $\zeta=0.6$ to 1.4 around a critical luminosity of about $L_{\rm{X}}=10^{35}-10^{36}$~erg~s$^{-1}$ \citep{2011MNRAS.414..677C}. XTE~J1752-23 also hinted at a transition where at lower radio/X-ray luminosity the ratios are similar to GX~339-4, but at higher luminosities the radio jet is apparently under-luminous if tested against the ``fundamental-plane''. Moreover, \citet{2014MNRAS.445..290G} performed a cluster and linear regression analysis on the hard and quiescent state of these XRBs (including a total of 24 systems). They found that a two cluster solution was favoured if the uncertainty of the luminosities is less than a factor of $\sim2$; however, if one includes data collected on Cygnus X-1 there is no significant evidence for a bimodal distribution of tracks.

Although our analysis for Swift\,J1753.5 clearly shows the source to be on the lower-branch of the $L_{\rm{radio}}/L_{\rm{X-ray}}$ relationship, it is not clear if the source  prefers a similar powerlaw relationship to that of GRS1915+105 or neutron stars (i.e. $\zeta\sim1.4$). Rather the scatter in the scaling relationship could be upper-branch variability albeit with a normalisation about an order of magnitude lower (to illustrate this in Figure~\ref{fig:luminosity} we show $\zeta\sim0.7$ at four different decades in blue dotted lines). Therefore, while we cannot rule out variations in the radiative efficiency, that leads a different scaling of the observed X-ray luminosity with accretion rate, we note other parameters (e.g. such as Doppler boosting) could also have an important role at these Eddington ratios.

Jet activity is not found in the soft X-ray states and suppression occurs when the temperature of the black body disk increases. However, it is not clear if the jet-quenching mechanism is linked to a {\it relative} change in the disk to hard powerlaw ratio or only an {\it absolute} drop in the hard powerlaw component. To help constraint this we use an XMM-Newton/NuSTAR measurement of the source studied in this paper during a different epoch. \citet{2016MNRAS.458.1636S} fitted a hard power law component of $\Gamma=1.79$ with a flux of $8.98^{+0.35}_{-0.24}\times10^{-11}$~erg~s$^{-1}$~cm$^{-2}$ (scaled to the 0.6-10 keV band) during a soft-state, which was about $\sim9\%$ of the total flux (the rest was dominated by the BB and an additional soft power-law component). If we {\it assume} the same fractional flux for the hard powerlaw during our JVLA soft-state observation, then the inferred powerlaw flux was $\sim6\times10^{-11}$~erg~s$^{-1}$~cm$^{-2}$. This flux is a factor of $\sim10$ less than the typical hard state level when the radio is around  $300 \mu$Jy. Thus if the radio flux scales just with the powerlaw component, we expect $L_{\rm{radio}}=300 / 10^{\zeta} \mu$Jy, which equates to 60 and 12 $\mu$Jy for $\zeta=0.7$ and 1.4 respectively. Although this is an estimate, it tentatively suggests that the hard powerlaw to radio flux does not scale with the standard $\zeta=0.7$ relationship. 

Soft-states have been seen to occur at accretion rates of $0.1-100\%~\dot{m}_{\rm{Edd}}$ \citep{2010MNRAS.403...61D}, therefore one may consider jet suppression to only occur at high accretion of up to (and possibly beyond) the Eddington accretion rate. However, \citet{2016MNRAS.458.1636S} measured the soft state to be at an unusually low accretion rate ($\sim0.6\%~\dot{m}_{\rm{Edd}}$), we can therefore {\it rule out} a requirement that the accretion rate has to be larger than $\sim1\%~\dot{m}_{\rm{Edd}}$ for jet quenching to occur.

The deep JVLA observations in the soft-state of Swift~1753.5 show the most accurate radio quenching of a transient BH XRB. Unlike the weak radio component seen in a soft-like state of Cygnus X-1, which may have been a failed state-transition \citep{2012MNRAS.419.3194R}, there is no radio emission associated with the soft-state of Swift~1753.5. Although we compare this result to the limit measured for 4U\,1957+11 \citep{2011ApJ...739L..19R}, it is important to note that that source has never been detected in the radio; thus the quenching factor for 4U\,1957+11 is an estimate based on assuming the hard state flux follows the upper branch of the disk-jet relationship, which is not known for 4U\,1957+11 and can vary widely between sources. However, since the jet flux is known for Swift J1753.5, our estimate can be much more accurate.

\section*{Acknowledgements}
We thank the EC for supporting ARP with a Marie Curie Inter-European fellowship under contract no. 2012-331977. The work was also supported by ERC grant 267697 ``4~PI~SKY: Extreme Astrophysics with Revolutionary Radio Telescopes''. P. G. thanks STFC (ST/J003697/2). DA acknowledges support from the Royal Society. MK acknowledges support from the European Union Seventh Framework Programme (FP7/2007-2013) under grant agreement no. 312789, Strong Gravity. We thank the Radio Astronomy Observatory (NRAO) for providing VLA data and the National Aeronautics and Space Administration (NASA) for providing Swift data. We also thank the staff of the Mullard Radio Astronomy Observatory, University of Cambridge, for providing AMI data. Finally we thank St\'{e}phane Corbel for sharing data on XRBs. 












\bsp	
\label{lastpage}
\end{document}